\documentclass[journal=jctc,manuscript=letter,maxauthors=20]{achemso}
\usepackage{achemso}
\setkeys{acs}{maxauthors = 999}
\usepackage[version=3]{mhchem}
\usepackage{titlesec}
\titleformat{\paragraph}
{\normalfont\normalsize\bfseries}{\theparagraph}{1em}{}
\titlespacing*{\paragraph}
{0pt}{3.25ex plus 1ex minus .2ex}{1.5ex plus .2ex}
\usepackage[utf8]{inputenc}  
\usepackage{graphicx}  
\usepackage{xcolor}  
\usepackage{amsmath}  
\usepackage{amsfonts}  
\usepackage{latexsym}  
\usepackage{amssymb}  
\usepackage{bm} 
\usepackage{setspace}
\usepackage{hyperref} 
\usepackage{caption}
\usepackage{mathtools}
\usepackage{subcaption}
\usepackage{tikz}
\usetikzlibrary{arrows, positioning}
\usepackage{qcircuit}
\usepackage{array}
\usepackage{tabularx}
\usepackage{multirow}
\usepackage[export]{adjustbox} 
\usepackage{physics}
\usepackage{csquotes}

\author{Davide Castaldo}
\affiliation{Department of Chemistry and Applied Biosciences, ETH Zurich, Vladimir-Prelog-Weg 2, 8093 Zurich, Switzerland}

\author{Markus Reiher}
\affiliation{Department of Chemistry and Applied Biosciences, ETH Zurich, Vladimir-Prelog-Weg 2, 8093 Zurich, Switzerland}
\email{mreiher@ethz.ch}

\title[]{Utility-scale quantum computational chemistry}

\date{March 19, 2026}

\begin{document}


\begin{abstract}
Chemistry and materials science are widely regarded as potential killer application fields for quantum hardware. While the dream of unlocking unprecedented simulation capabilities remains compelling, quantum algorithm development must adapt to the evolving constraints of the emerging quantum hardware in order to accomplish any advantage for the computational chemistry practice. At the same time, the continuous advancement of classical wavefunction-theory methods narrows the window for a broad quantum advantage. Here, we explore potential benefits of quantum computation from the broader perspective of utility-scale applications. We argue that quantum algorithms need not only enable accurate calculations for a few challenging, that is strongly correlated, molecular structures, that might be hard to describe with traditional methods. Instead, they must also support the practical integration of quantum-accelerated computations into high-throughput pipelines for routine calculations on arbitrary molecules, ultimately delivering a tangible value to society.
\end{abstract}

\section{Introduction}
The past decade has seen a rush into quantum computation applications\cite{biamonte2017quantum, mcardle2020quantum, bayerstadler2021industry, portmann2022security, herman2023quantum, baiardi2023quantum}, with many focusing on chemical and materials science problems as these fields are believed to benefit from this new computing paradigm significantly\cite{reiher2017elucidating, goings2022reliably, hoefler2023disentangling, alexeev2025perspective}. 
How to define an associated quantum advantage in a meaningful way has been the subject of ongoing considerations\cite{eisert2025mind, babbush2025grand}.
So far, the demonstration of a quantum advantage for quantum chemistry on near-term quantum computers mostly focused on strong electron correlation examples from the molecular world, where traditional quantum chemical methods face severe challenges\cite{hastings2007quasi, eisert2010colloquium}. A prime example has been the mechanism of nitrogenase\cite{reiher2017elucidating} which remains to be a conundrum and one of the Holy Grails of research in chemistry\cite{einsle2020structural, jasniewski2020reactivity, qing2020recent}. Many related problems can easily be formulated, such as elucidation of the mode of action of the manganese cluster in photosystem II\cite{cox2020current, greife2023electron, bhowmick2023structural}, of the iron-sulfur cluster in FeFe hydrogenases\cite{fontecilla2007structure, vogiatzis2018computational, trncik2023iron}, or of Rieske dioxygenases\cite{runda2023rieske, liu2022design}. However, challenging electronic structures are also found in other fields such as hetereogenous catalysis\cite{steiner2022autonomous}.

As long as quantum algorithms are not competitive with traditional computations—due to a limited number of logical qubits and limited execution times of current and mid-term quantum computers—and as long as quantum computers are expensive to build and operate, algorithms offering exponential speedups will continue to be a central focus of research\cite{shor1999polynomial, childs2003exponential, gilyen2019quantum, babbush2023exponential}. These limitations forced research to focus on domains where classical methods fail.

However, this point of view is too narrow for potentially useful applications in molecular and materials science as there exist hardly any chemical problems where a single high-accuracy calculation will make a key difference. Instead, many computational (and experimental) results play in concert to create a consistent picture and solve a problem\cite{von2021quantum, goings2022reliably, capone2024vision}. Hence, many, if not most, of the calculations dedicated to solving a chemical puzzle will have to deal with weakly correlated (single-reference) structures, for which coupled cluster models have been established as a remarkably successful approach in the last three decades\cite{ raghavachari1989fifth, bartlett2007coupled, bartlett2024perspective}. 

Although traditional computational chemistry results may suffer from sizable errors, reasonable chemical conclusions may still be drawn from them\cite{jensen2017introduction, koch2015chemist}. In fact, this latter observation is one reason for the success of density functional theory in computational chemistry: despite some dramatic failures (see Refs. \citenum{reiher2002theoretical, dreuw2004failure,  plata2015case, cao2019extremely} for some examples), it often delivers useful results with sufficiently small errors\cite{kirklin2015open, mardirossian2017thirty, bursch2022best}. Still, the error of a specific calculation is unknown, and therefore one usually attempts to infer its magnitude from a combination of known results for the system under study and from benchmark data for similar systems. Note that also modern classical wavefunction methods\cite{motta2017towards, wu2024variational} suffer from unknown system-dependent errors\cite{reiher2022molecule}. By contrast, quantum phase estimation (QPE) based approaches can guarantee an arbitrarily small $\epsilon$-accurate energy estimate (within a given one-particle basis), given a sufficient amount of quantum resources\cite{nielsen2010quantum}.

In view of this current situation in computational chemistry, we are led to a different paradigm for quantum computations of general value in chemistry and materials science. This is one that demands utility-scale (fault-tolerant) quantum computations, done in parallel and routinely. Rather than solely focusing on a quantum advantage, the new paradigm would imply a far broader perspective for quantum computations.

In this Perspective, we shall therefore address key questions regarding general applicability of quantum computation: what is its value for general computational chemistry? Is it meaningful to focus on strongly correlated molecules to be simulated on a quantum computer? How can we efficiently integrate current Quantum-Processing-Unit-(QPU)-accelerated routines into multiscale simulations? 

We need to ponder these aspects carefully because we live in a world of limited resources being exploited beyond sustainable limits\,\cite{masson2021climate, richardson2023earth}. It is therefore imperative to assess whether, from both an environmental and an economic perspective, the effort we are making to develop and run these machines is reasonable\,\cite{auffeves2022quantum}. We will come back to this point in our conclusions.

This work is organized as follows: Section \ref{qcs} discusses the current quantum computing stack (i.e., the hierarchical conceptualization of the quantum computational workflow from the user interface down to the physical instructions on the hardware), necessary to break down the analysis of utility-scale quantum computation for chemistry into different parts. Subsequently Section\,\ref{targets} elaborates on the requirements and conditions under which an ideal realization of the quantum computing stack could provide practical advantages over emerging modern classical wavefunction approaches. Finally, we summarize our discussion and point towards future developments of quantum algorithms for chemistry.

\section{The quantum computing stack}
\label{qcs}

How does the path for useful quantum computation look like? This question is concerning scientists in academia and the private sector all over the world\cite{brooks2023race}. At the outset of the computing era, programmers were forced to develop their codes in machine language, making each software development incredibly challenging and virtually impossible to directly port on another machine. A major shift came with the introduction of compilers and Intermediate Representations (IRs)\cite{appel1998modern}. These new tools enabled the creation of what is nowadays called the computing stack: a hierarchical conceptualization of any computational workflow from the user interface down to the physical instructions on the hardware.
With the advent of quantum processors\cite{monroe1995demonstration}, a natural generalization of this concept to quantum computers is inevitable. The quantum computing stack\cite{beverland2022assessing} can be thought of six different layers: 

(i) At the bottom of the stack we have the hardware, i.e. the physical substrate hosting the computational device. Several proposals for this substrate are currently under consideration among the scientific community spanning different states of matter ranging from light, molecules and solid states materials\cite{de2021materials} to more exotic topological phases\,\cite{aasen2025roadmap}; the variety and open competition among the different physical substrates reflects that no clear advantage has been established yet for any of the main proposals. 
It is well known that superconducting qubits\cite{bravyi2022future} outperform other platforms in terms of repetition rate of a given circuit but, at the same time, they lag behind neutral atoms and ion-based platforms in terms of connectivity\cite{lekitsch2017blueprint, scholl2021quantum, bluvstein2024logical}. To this end, we note that a superconducting-based architecture with an effective all-to-all connectivity was recently proposed\cite{renger2025superconducting, vigneau2025quantum}. At the same time, photonic based platforms can leverage the expertise acquired in the field of integrated photonics with photon generations being cheap, fast, and able to operate at room temperature\cite{bourassa2021blueprint, bartolucci2023fusion}.  

(ii) Second, we have the physical qubits, that is, the particular two-level quantum states within the hardware eigenspectrum used to actually perform computation. This is followed by (iii) the quantum error correcting code. The latter can be thought as a mapping $\mathcal{C}$ between a quantum circuit $Q$ defined on the set of $N$ physical qubits to a quantum circuit $Q'$ defined on the set of $N'$ logical qubits ($\mathcal{C}(Q): Q \xrightarrow{} Q'$). Such a mapping typically requires the encoding of the logical information in a redundant physical space and the use of error-correction and detection algorithms\cite{lidar2013quantum} to avoid information loss due to the presence of an external environment and imperfect gate implementations. (iv) The abstract objects resulting from the error correcting procedure are defined as logical qubits, the set of operations, acting on the logical qubit, which define the universal model of computation is referred to as Instruction Set Architecture (ISA). (v) The Quantum Intermediate Representations (QIR) comprise a middle layer of operations which enable a compiler friendly and hardware-agnostic procedure to translate from the source code to the machine language. (vi) The quantum algorithms are high-level strategies to solve a computational task on the QPU.

With this picture in mind, it is now easier to understand that for developing a quantum algorithm it is mandatory to think of the overhead incurred by compiling our algorithm down through the whole quantum computing stack. It is worth considering in more detail this overhead. If we consider the problem of connectivity, one must account for the fact that the $N$ physical qubits are separated in space after the original quantum circuit has been translated in terms of the ISA. To implement some operations, depending on the architecture at hand, additional gates are needed which can account for up to $\mathcal{O}(N)$ additional Control-NOT gates (CNOTs) when only a linear nearest-neighbor connectivity is allowed\cite{dreier2025connectivity}. Subsequently, one must take into account the computational overhead due to the error correcting code. 
As noted before, quantum error correction (QEC) involves the mapping of the initial logical circuit into a redundant space which is used to store the information (code space) and to on-line detect the presence of an error by syndrome measurements and decoding by classical algorithms\cite{bausch2024learning, piveteau2024tensor, higgott2025sparse} . Diving into the technical details of error-correction theory is beyond the scope of this perspective but, for the purpose of our discussion, it is important to consider the main factors that induce an overhead due to error correction.

Particularly, in order to implement an error corrected circuit we need two ingredients: (i) a quantum memory (i.e. the ability to store a quantum state idle) and (ii) a quantum logic (i.e. the ability to perform fault-tolerant logical operations). Both the logical operations and the logical states are defined by the quantum error correcting code at hand.
 
Each logical qubit is mapped into a set of \textit{data} qubits and each logical operation acts on them accordingly. The latter point implies an additional overhead because not all the gates can be implemented in a fault-tolerant manner with a trivial mapping from the logical space to the physical space\cite{eastin2009restrictions}. Particularly, a typical convention in the QEC literature is to distinguish between \textit{transversal} and \textit{non-transversal} gates. We can, intuitively, define as transversal each gate that, when mapped into the QEC circuit, is just the independent application of the logical gate on each physical qubit. As an example of transversal operation, we consider a 3-qubit encoding of a logical qubit ($|0_L\rangle\xrightarrow{}|000\rangle , |1_L\rangle\xrightarrow{}|111\rangle$); if we want to apply the $X_L$ gate (i.e. the Pauli \textit{X} on a logical qubit $X_L|0_L\rangle = |1_L\rangle$), we then have to apply independently the \textit{X} gate on each of the data qubits, $X_L|0_L\rangle = XXX|000\rangle=|1_L\rangle$. Inevitably, not all operations will satisfy this property. For those cases, we will need additional ancilla (i.e., auxiliary) qubits to consume and measure. When the non-transversal gate used to achieve a universal gate set is the $T$ gate, these additional sets of qubits are called $T$ factories.

So far, we have introduced two sets of qubits: one used to encode the logical qubits and another one used to perform logical operations. To construct a quantum memory, we further require an additional set of qubits to detect and correct, in real-time, errors due to decoherence. These additional qubits are referred to as syndrome qubits. During an error correction cycle, a set of CZ gates is applied between the data qubits and the syndrome qubits, which are then measured. The number of measurements performed is related to the code distance $d$, i.e., the maximum number of correctable errors in a given QEC code. The resulting syndrome measurements form a dataset used to infer, via a classical decoding algorithm, the inverse operator $\mathcal{E}^{-1}$ to be applied in order to correct the error $\mathcal{E}$. Finally, to achieve full fault-tolerance, we must also account for the possibility that syndrome measurements themselves could be faulty. In such cases, an additional set of ancilla qubits is required.

All the current proposals for quantum error correction aim to minimize the overall time required for an error correction cycle and the total number of additional qubits needed to embed the logical circuit. Particularly, the major contribution originates from the computational overhead required for the synthesis of the logical $T$ gates\cite{fowler2012surface, li2015magic, lee2024low}, which is the basis of the current estimates of fault-tolerant quantum simulations for chemistry\cite{beverland2022assessing, delgado2022simulating, kim2022fault, blunt2022perspective}. Note the advancement demonstrated by magic state cultivation\,\cite{gidney2024magic, vaknin2025magic}.

Since the goal of this Perspective is to contextualize the field of quantum computational chemistry through the lens of utility, the above discussion on the quantum computing stack shall make clear that many different factors kick in when it comes to determining the actual cost (i.e. the utility) of a quantum algorithm. The key message that we want to convey so far is that the inherent nature of quantum computers calls for the implementation of error correcting code but, at the same time, their cost calls for a serious reflection on the conditions under which this is the best option and when other form of denoising are more suitable.

In the next paragraph, we will make use of a very recent model for scalability of quantum processors\cite{katabarwa2024early} to qualitatively delineate different regimes of compilation that one may want to explore when developing quantum algorithms for quantum chemistry. 

Before proceeding with our discussion we note that scaling quantum hardware presents distinct challenges across different architectures. Ion-trap systems struggle with connectivity at scale, as shuttling ions between trap zones becomes prohibitively slow. While entangling ions across separate traps using photonic links offers a potential solution, its implementation remains technically demanding\cite{main2025distributed, mordini2025multizone}. Superconducting qubits face fundamental wiring constraints, because the sheer density of control lines is becoming unmanageable as systems grow. This and other challenges have been documented in detail for superconducting platforms in Ref.\,\citenum{mohseni2024build}. Photonic approaches, based on Gottesman-Kitaev-Preskill encoding\cite{aghaee2025scaling} or dual-rail schemes\cite{psiquantum2025manufacturable}, are primarily bottlenecked by photon loss, requiring dramatic improvements in transmission and detection efficiency. Across all platforms, classical control electronics and quantum operating systems must also advance significantly to manage the complexity of large-scale quantum operations in real time\cite{mohseni2024build}. Particularly, the last issue implies that even scaling up the number of qubits and their quality would not necessarily imply advancing the clock rate of the quantum processor. Hence, we must assume an optimistic perspective on these general hardware challenges for the discussion in the following sections.

\subsection{Different regimes of compilation}
\label{paradigms-of-compilation}

If we assume that each operation on a quantum computer fails with a probability $p_{\text{gate}}$, and that all such failure events are independent, then the probability that a quantum circuit is executed without any errors is given by $p_{success} = (1 - p_{\text{gate}})^{\#_{\text{gates}}}$. Since the fidelity $F = |\langle\Psi_{exact}|\Psi\rangle|^2$ is affected by the probability of successfully running the circuit, as a first approximation, we may assume that the two are identical: 

\begin{equation}
\label{fidelity}
     F = p_{success} = (1 - p_{\text{gate}})^{\#_{\text{gates}}} \, \, .
\end{equation}

If a full-fledged error-corrected quantum computer is realized, it would imply that error-correcting codes are capable of counteracting this exponential decay in fidelity. In fact, this is formalized in the celebrated threshold theorems\cite{aharonov1997fault, knill1998resilient}, which guarantee that increasing the code distance (i.e., the number of correctable errors) allows an error-correcting code to exponentially suppress logical error rates, provided that the physical error rate of each gate is below a code-dependent threshold value $p_{\mathrm{th}}$. An experimental demonstration of this principle has been recently reported in Refs.\,\citenum{eickbusch2024demonstrating, acharya2024quantum} where the authors employ the surface code at different code distances $d$, using $d^2$ physical qubits and $d^2 - 1$ syndrome measurements to encode a single logical qubit.

\begin{figure}[htbp!]
    \centering
    \includegraphics[width=0.7\linewidth]{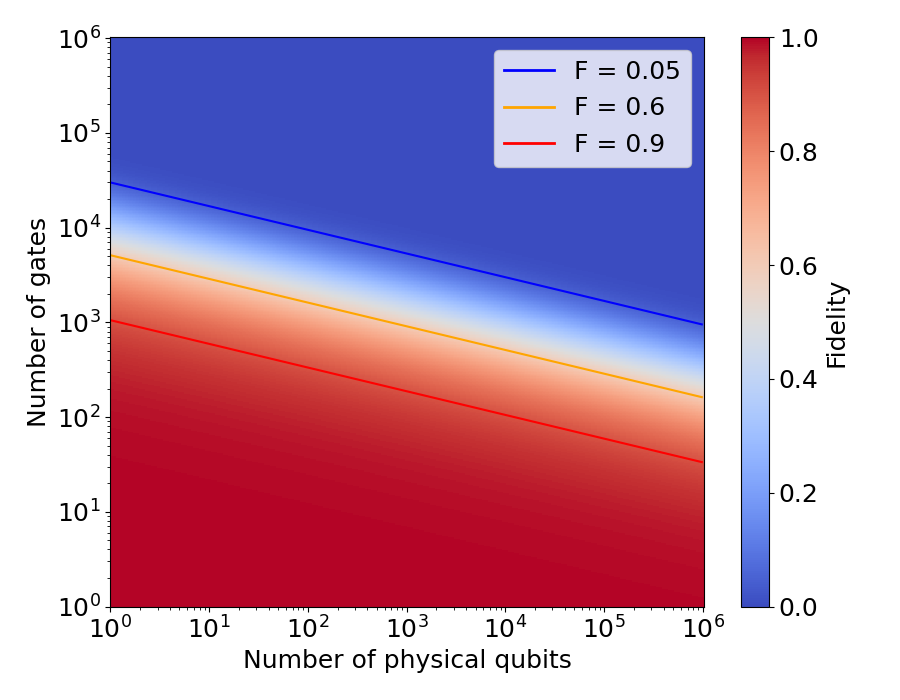}
    \caption{Quantum circuit fidelity \( F \) modeled according to the scalability framework proposed in Ref.\,\citenum{katabarwa2024early}. Fidelity decreases exponentially with the number of gates, following Eq.\,(\ref{fidelity}). The gate error probability \( p_{\text{gate}} \) decreases with the number of physical qubits \( Q \) and is given by \( p_{\text{gate}} = p_0 Q^{1/s} \), where \( s \) and \( p_0 \) are the hardware-dependent scalability factor and average error rate, respectively. Here we set \( s = 4 \) and \( p_0 = 0.0001 \).
}
    \label{circuit-fidelity}
\end{figure}

In this section, we aim to use Eq.~(\ref{fidelity}) to suggest that the development of quantum algorithms—alongside advances in hardware—calls for different compilation paradigms depending on the size/quality of the available QPU. In Fig.~\ref{circuit-fidelity}, we show the fidelity of circuit execution as a function of the number of physical qubits and the number of gates. To compute $F$ we used Eq.\,(\ref{fidelity}) also accounting for qubit degradation that can arise when scaling up quantum hardware architectures, as previously discussed in Ref.\,\citenum{katabarwa2024early}. For the purposes of this discussion, we have also plotted three (solid) contour lines corresponding to decreasing values of $F$. The values of $F$, which determine the position of these lines, depend on the specific hardware and, in particular, on that hardware feature which mostly affects the denoising scheme considered in our compilation. Error mitigation speed is mostly affected by the repetition rate of a quantum processor, while the error correction and detection are strongly dependent on the connectivity of the architecture which enforces constraints on the code that can be hosted on a given QPU. Due to the exponential decay of $F$, these contours suggest the existence of four distinct operational regimes for a quantum computer.

As discussed above, implementing an error correction protocol entails a substantial overhead in computational resources, including qubits, gates, and measurements. Hence, it is reasonable to consider alternative strategies that have been proposed in the literature, such as Quantum Error Mitigation\cite{cai2023quantum} (QEM) and Quantum Error Detection\cite{knill2004fault, knill2004fault2} (QED), where we trade a lower computational overhead, in terms of additional physical qubits and gates, for worse performance, in terms of error reduction.

QEM is generally considered in the context of estimating the expectation value
of an observable using a quantum circuit by reducing the noise-induced bias\cite{kandala2019error, kim2023scalable}. This is done through post-processing of the measurement
outcomes from an ensemble of circuit implementations. Consequently, most error mitigation techniques require additional sampling that increases with the noise
in the circuit; unfortunately, this results in an asymptotically unfavorable scaling\cite{cai2023quantum}. By contrast, QED uses, in the same spirit of the syndrome measurements discussed above, measurements on ancilla qubits to detect the presence of an error. Within this scheme, each measurement flagging an error is discarded for the purpose of the calculation\cite{knill2004fault, knill2004fault2}. Giving that QEM typically requires no additional physical qubits and QED only requires a few additional qubits per computational qubit, we can foresee four different compilation regimes matching the continuous evolution of quantum hardware. In the following, we discuss each of these regimes.

\paragraph*{Full QEM compilation}

In this first setting, the number of available physical qubits is very limited, and no additional resources can be allocated to the implementation of error reduction protocols. To this end, the only viable option is to use QEM techniques to reduce the noise-induced bias in the computation. The algorithms that can be implemented on this type of devices typically involve very shallow circuits, where quantum coherence cannot be significantly degraded by the external environment.

In Fig.~\ref{circuit-fidelity}, we observe that the amount of resources available before hitting the $F = 0.9$ line corresponds to approximately 500 gates and 500 physical qubits. These numbers imply that one must optimize the total cost in order to reduce circuit depth, and that each QEM computation typically consists of the execution of very many different circuits. Consequently, the most commonly adopted quantum algorithms in this framework are variational which do not require ancilla qubits~\cite{cerezo2021variational}.

Achieving a practical quantum advantage in this regime appears very unlikely. Although the number of qubits might, in principle, allow one to work with wavefunctions beyond the reach of current state-of-the-art supercomputers, the number of circuit executions required by these algorithms to compute energies and other molecular observables easily becomes too large, because it scales as $\mathcal{O}(||A||_1/\epsilon^2)$, where $||A||_1$ is the 1-norm of the measured operator $\hat{A}$ and $\epsilon$ is the desired precision\cite{gonthier2022measurements}. We note that, to mitigate this unfavorable scaling, significant efforts have been made\,\cite{patel2025quantum, lee2025efficient}. 

Sample-based quantum diagonalization methods\,\cite{robledo2024chemistry, nakagawa2024adapt, barison2025quantum} work well in this regime. In these approaches, the quantum computer is solely used to sample electronic configurations, which are then employed to classically solve an eigenvalue problem within a subspace relevant to the simulated system, but the competition with existing classical methods makes it difficult to clearly identify a useful quantum advantage\,\cite{reinholdt2025exposing}.

\paragraph*{Mixed QEM/QED compilation}
\label{qem-qed}

By increasing the number of gates and/or qubits involved in the computation, the resulting noise becomes such that the exponential overhead introduced by error mitigation techniques will render our efforts to extract value from the computation futile. However, with access to a sufficient number of qubits we can allocate some of them for implementing error detection routines on selected operations, while still targeting computationally challenging simulations for classical computers. Specifically, in this compilation regime, we consider QPUs with around $10^4$ physical qubits and circuits comprising at most $10^3$ gates. We point to the work of Ref.\,\citenum{zhong2025combining} that very recently explored the ground state energy simulation of $\text{H}_2$ under this compilation regime.

In this spirit, algorithms for this regime of compilation still require the execution of many circuits per computation, but we can now target slightly deeper circuits. As in previous settings, the hardware dependent compiler will aim to minimize both the circuit depth and the number of CNOT gates.  It is important to highlight that in this regime we can now consider algorithms that have been theoretically proven to satisfy Heisenberg scaling of the runtime $T = \mathcal{O}(\epsilon^{-1})$, i.e. saturating the bounds of quantum information\cite{giovannetti2006quantum}. This can be accomplished because part of the denoising is carried out by classical subroutines\cite{ding2024quantum, yi2024quantum, ni2023low, dutkiewicz2024error, castaldo2025heisenberg, stroschein2025ground}. However, the success of this extraction remains strongly dependent on the level of noise present in the quantum data. Hence, the most appropriate algorithm of this kind for this regime are likely single-ancilla phase estimation algorithms~\cite{dutkiewicz2022heisenberg}. The key idea behind these methods is to decompose the quantum algorithm, such as the quantum phase estimation\cite{nielsen2010quantum} for energy measurements, into smaller circuits, each responsible for independently sampling a point of the molecular autocorrelation function. Based on these samples, an external classical subroutine extracts the frequencies of interest. This approach can also be extended to the extraction of general molecular observables\cite{zhang2022computing}. We point out that these algoritms, at least considered in their abstract formulation, could be competitive choices with respect to standard QPE, even when compiled into machines with a greater number of physical qubits.

Due to the limitations on circuit depth, the quantum simulation routines best suited for this compilation regime are those based on low-order Trotter formulas, which have recently benefited from several advancements~\cite{gunther2025phase, wan2022randomized, hagan2023composite, kiumi2024te}. The advantages of developing resource-aware algorithms were recently demonstrated by Toshio\,\textit{et al.}~\cite{toshio2025practical}, who proposed a framework that incorporates most of the features discussed in this section. Their comprehensive protocol estimates selected eigenvalues of an $8 \times 8$-site Hubbard model (whose exact solution is currently beyond classical computational capabilities) with an estimated serial execution time of approximately 10 days. Other notable contributions in this direction are described in Refs.~\citenum{yamamoto2024demonstrating, blunt2023statistical}.

\paragraph*{Mixed QED/QEC compilation}

When approaching the threshold of $10^5$ physical qubits, we can envisage a mixed compilation regime in which the error correction gadgeting procedure is applied only to a subset of the operations, while error detection protocols are implemented for the remainder. Due to the computational overhead required by T-gate factories, in this regime the compilation is aimed not at reducing the circuit depth but at minimizing the T-gate count; see Ref.\,\citenum{vandaele2024lower} for an example. In the same spirit, we highlight the very recent work of DalFavero and LaRose\,\cite{dalfavero2025error} who investigated the interplay of QEM and QEC compilation.

Considering that only a moderate number of gates can be error corrected, alternative strategies to reduce the qubit overhead such as QEM-based magic state factories could also be beneficial\,\cite{luthra2025unlocking}. Finally, from a more algorithmic perspective, it is known that QPE-like approaches rely on a high overlap between the input state and the target states\,\cite{erakovic2025high, fomichev2024initial} to reduce the total number of circuit executions. In the same spirit, strategies based on transversal gates-only circuits for preparing high-overlap initial states are a promising direction to reduce costs, even when QEC is only partially done\,\cite{anand2024stabilizer, sun2024stabilizer}.

\paragraph*{Full QEC compilation}

All computational tasks requiring more than \(10^5\) gates are likely to be addressed only by fully quantum error-corrected (QEC) machines. Algorithms designed for these machines to achieve utility-scale computation must fulfill several requirements: (i) algorithmic primitives (such as standard linear algebra operations) are efficiently compatible with native gates and their corresponding QEC gadgetization, (ii) robustness to poor initial state preparation, and (iii) high parallelizability.

It is also important to note that, at this level, quantum simulations should focus on those algorithms offering the best asymptotic scaling (with respect to system size) in terms of T-gates. For this reason, the quantum simulation routines, which will have the greatest impact these machines become available, are those based on quantum signal processing techniques~\cite{low2017optimal, berry2024doubling} or on decompositions of the time-evolution operator as a linear combination of unitaries~\cite{babbush2018encoding}. Currently, substantial effort~\cite{low2025fast, patel2025global} in this field is devoted to reducing the computational cost of these routines, which heavily rely on block-encoding subroutines~\cite{gilyen2019quantum, liu2025block} required to implement non-unitary operations.

\paragraph*{Conclusions on compilation regimes}

We emphasize that, to date, no clear demonstration of a useful quantum advantage has been achieved yet. The complexity of the quantum stack we have presented is reflected in a continuum of possible compilation strategies, which we have categorized into four distinct regimes. The key message is that adopting a hardware-dependent algorithmic design perspective can lead to practical utility for quantum computing in chemistry without necessarily requiring hardware specifications that match the full QEC regime. In particular, the hardware-algorithm co-design discussed above could provide a practical demonstration of the advantage of quantum algorithms over classical methods in routine calculations within a regime corresponding to the mixed QED/QEC compilation (i.e. with $\approx 500-1000 $ logical qubits).

To understand why this type of compilation could be advantageous, we consider a circuit in which the probability of an error occurring in each operation is $p_{gate}$. It can be shown that the sampling overhead of QEM/QED required for circuits with this type of noise is modest (compared to the exponential overhead arising for arbitrary circuit depth) provided that the circuit size is of order $p_{gate}^{-1}$\cite{tsubouchi2023universal, martiel2025low}. In this case, the idea of the mixed compilation paradigm is that QEC protocols do not necessarily need to reduce logical error rates to the level of the standard floating-point errors of classical computers $\epsilon \approx 10^{-16}$, but rather to achieve only partial error correction such that $p_{gate}^{pc} < \epsilon$, with $(p_{gate}^{pc})^{-1} \approx \mathcal{G}$, where $\mathcal{G}$ denotes the size of the circuit to be executed. Very recently, similar circuit size estimates for utility scale quantum computation have been made exploring the impact of a QEM/QEC compilation\,\cite{aharonov2025importance, zimboras2025myths}.

\section{Target Applications in Molecular and Materials Science}
\label{targets}

Quantum computers are expensive to develop and so there must be some way to create significant benefits to make up for the investments. Considering electronic structure simulations, the question is therefore where traditional computational quantum chemistry and materials science is currently employed to advance knowledge and make discoveries of potential industrial value. This is the case for understanding, design, and improvement of materials and catalytic processes, and potentially of pharmacological events. With this in mind, a main task driving the quantum computing community working on electronic structure has been the identification of target systems to demonstrate a quantum advantage. 

An $N$-electron wavefunction expressed in a basis set of $M$ spin-orbitals lives in a $\binom{N}{M}$-dimensional Hilbert space, but what eventually matters is how many electronic configurations actually contribute to the accurate description of a system. To understand this problem, well-established categorization schemes for the electronic structures of molecules and materials in terms of electron correlation can be utilized\,\cite{langhoff1974configuration}. Standard categorization schemes are based on the number of electronic configurations with large amplitude in the $N$-electron wavefunction and the number of orbitals to be included in the calculation. Accordingly, one splits the total electronic correlation problem into static and dynamic correlation\,\cite{ruedenberg1982atoms,diercksen1983electron,bofill1989unrestricted,lee1989diagnostic, jiang2012multireference}.

The separation of the total electron correlation problem into static and dynamic correlation implies two subsets of orbitals (apart from the lower lying core orbitals): the active orbital space contains the strongly correlated ('active') orbitals, whereas the typically much larger set of weakly correlated orbitals gives rise to the dynamic correlation. The active orbitals pose a challenge to traditional electronic structure methods because of the resulting multi-configurational nature of the electronic wave function. Exact diagonalization within the active space followed by the addition of the dynamic correlation through perturbation theory is the standard procedure to deal with the different types of orbitals (giving rise to complete active space approaches with subsequent multi-reference perturbation theory). As long as future fault-tolerant quantum computers cannot represent electronic states on a few thousand logical qubits, the separation of the orbital space into active orbitals and their complement will be a hardware-imposed necessity (simply because of the limited number of logical qubits available for state preparation and phase estimation). Hence, the consideration of dynamic correlation will be a mandatory separate step for quantitative predictions, which may be accomplished by quantum variants of multi-reference perturbation theory (as in Refs.\citenum{gunther2024more, mitarai2023perturbation}) or by dressing the Hamiltonian in such a way that dynamic correlations are folded into the Hamiltonian prior to quantum simulation (see, e.g., Refs. \citenum{hedegaard2015density, bauman2019quantum,szenes2024striking}).  

A remaining issue is the identification of the active orbitals, for which concepts of orbital entanglement\,\cite{legeza2003optimizing, legeza2004quantum} allow one to accurately discriminate between the two types of electron correlation\cite{boguslawski2012entanglement}. Along these lines, we developed\,\cite{stein2016automated, stein2016delicate, stein2019autocas} an automated protocol, the so-called autoCAS protocol, to determine the relevant active space for the description of a molecule. 
This protocol can also be applied to categorize the electronic structure of a given molecule and to hint at the number of determinants to include in QPE state preparation\cite{morchen2024classification}. In the categorization scheme introduced in Ref.\,\citenum{morchen2024classification}, molecules placed into class 0 can be qualitatively described by a single electronic configuration and therefore feature only dynamic correlations. By contrast, multi-configurational electronic structures are grouped into two classes\,\cite{morchen2024classification}, class 1 and class 2, where the latter class comprises those cases that require a large fraction of the full valence orbital space as active space. 
Class-2 cases are very hard to handle on classical computers and therefore considered as ultimate targets for quantum computation.

By contrast,
class-1 molecules are associated with comparatively small active spaces. In chemistry, they represent the majority of the multi-configurational cases (for instance, when considering transition states in chemical reaction space, mononuclear open-shell transition-metal complexes, or excited states in visible-light photochemistry). Accordingly, many well-established traditional multi-reference approaches have been developed for their quantum chemical description. Accordingly, the importance of the weakly correlated class-0 and the moderately correlated class-1 cases in chemistry points to the need for utility scale quantum computation run routinely for standard static correlation problems and even for weakly correlated molecules. 

Still, we may point to application areas, where typical class-2 molecules can be found. In view of the notorious nitrogenase mechanism example\cite{reiher2017elucidating}, it is obvious that the mode of action of metalloproteins with multi-nuclear transition metal clusters (such as FeFe hydrogenase or the manganese cluster in photosystem II mentioned in the Introduction) is a source for class-2 examples. Since these clusters may be understood as fragments of solids, such as chalcogenides and oxides, class-2 systems are likely to be found frequently among intermediates of heterogeneous catalysis. If we consider a rough surface of a metal, a metal-chalcogenide, a metal-oxide, or of an alloy, high structure variability and unsaturated valences can produce strong electron correlation effects for substrate-binding surface atoms. Such situations can be dealt with by applying projection embedding approaches\cite{vorwerk2022quantum, rossmannek2023quantum, battaglia2024general, chen2025advances, weisburn2025multiscale, bensberg2025hierarchical} to define a sufficiently large structural model of such a piece of a surface that is then likely to have class-2 character. 

While ab initio predictions in general demand accurate calculated data, this becomes most obvious in the context of microkinetic modelling where energy errors enter rate expressions through an error-amplifying exponential~\cite{proppe2016uncertainty} of the rate constant. In all such applications, quantum computation (through phase estimation) is expected to deliver energies of controllable accuracy in a given orbital space.

Next, we should put these application cases into the broader perspective of future computational chemistry and materials science. We will soon see a shift away from DFT as the standard workhorse of computational campaigns to (foundation) machine-learned interatomic potentials (MLIPs)\,\cite{behler2007generalized, bartok2017machine, batatia2022mace, kalita2025machine, lysogorskiy2026graph} . This paradigm shift has significant implications for quantum computation in molecular and materials science. Quantum computations will then serve the purpose of training data production and validation. Hence, high-quality training data obtained from QPU-accelerated calculations will allow for the development of more accurate machine learning procedures. 

We may expect future quantum computation to become a true competitor to classical approaches only if it can be carried out routinely also for weakly correlated problems within the resource limits set by the best traditional approaches (that is, within the hardware resources and runtimes needed for accurate coupled cluster calculations). 
If no active space needs to be defined (as for class-0 molecules) the benchmark method remains CCSD(T)\cite{raghavachari1989fifth}. This is why quantum computational chemistry must compare to the runtime needed to reach chemical accuracy with CCSD(T). In the spirit of this broader perspective on quantum computation, we point to the work of Chen and Chan\,\cite{chen2025framework} who have begun exploring a theoretical framework for general quantum speedups in quantum chemistry.

To demonstrate a gain in time and utility for large-scale quantum chemistry simulations, we can consider two possible strategies. The first consists of directly comparing the time required by the two methods, CCSD(T) and QPE-type approaches, to obtain a result with an error equal to $\epsilon$ relative to the exact energy in a given orbital basis set. The second is a more theoretical approach, in which we may estimate the speedup $\Lambda$ by calculating the ratio of the error bounds for the two methods as a function of the orbital space $\Lambda = \mathcal{O}\!\left(\frac{\varepsilon_{\mathrm{QPE}}(N)}{\varepsilon_{\mathrm{CCSD(T)}}(N)}\right)$, for a fixed amount of computational resources (classical and quantum). 

Considering the first approach, however, we currently cannot perform QPE calculations and directly measure the elapsed time. The predominant approach in the literature has therefore been making assumptions about the specifications of the available hardware, its architecture, and the error-correction protocol aiming to identify the hardware specifications required to guarantee a quantum speedup\cite{harrigan2024expressing, gratsea2025achieving}. Although this introduces a certain degree of arbitrariness in the estimation of the elapsed time, it nevertheless allows for comparisons of computational costs that would otherwise be inaccessible. 

Considering the potential roadblocks that different platforms may encounter in their scalability over the coming years (see Sec.\,\ref{qcs}), it appears likely that early quantum computer prototypes will offer greater availability of logical qubits rather than large circuit depths. Therefore, detailed studies of runtime estimates for quantum algorithms based on single ancillas are a priority at present as such algorithms are naturally embarrassingly parallel. Moreover, by construction, they require shallower circuits and reduced connectivity, since they perform conditional evolutions on only one qubit at a time. Assuming a quantum computer in which the width (i.e., the number of logical qubits) is not a computational bottleneck, but the depth (i.e., the number of logical operations) is, this allows us to also mitigate the issue of clock speed in future quantum processors. Such processors are likely to operate very slowly in their early stages, in view of current error-correction protocols and the challenges associated with real-time calibration. However, such an approach will also be more architecture-friendly, especially in scenarios where achieving a high qubit count involves connecting multiple QPUs—each containing hundreds or thousands of qubits~\cite{aghaee2025scaling, gambetta2020ibm}.

If, conversely, we focus on the second approach, it becomes apparent that the primary limitation is the absence, to date, of rigorous error bounds for coupled-cluster methods that can be directly compared with those available for QPE-type algorithms. One of the central open problems in modern computational chemistry is therefore how to derive rigorously such estimates, not only for coupled-cluster methods, but also for the full spectrum of techniques developed over recent decades\cite{reiher2022molecule}.

\section{Conclusions}
\label{conclusions}

Building a quantum computer is a tremendous challenge. Deploying scalable, useful, and affordable quantum hardware will be crucial, but its availability will depend on many factors.
A truly reliable and useful quantum machine would feature all-to-all qubit connectivity, low error rates on native operations---low enough to enable quantum error correction with manageable overhead---fast and accurate readout, scalability without qubit degradation, and relatively low production costs. Since these features are not likely to be accomplished altogether in the near future, we must take them as criteria to critically consider quantum algorithm development.

In this work, we therefore elaborated on what a quantum computer should be able to target in order to be useful for the computational chemistry practice. One goal must be a computational advantage with respect to methods that already provide accurate results (such as CCSD(T)) in calculations on classical computers. Next, it must be possible to perform such quantum calculations routinely, rather than targeting single highlight results for very difficult problems.  

In essence, we should aim not only at the celebrated \textit{chemical accuracy} in single-point calculations, but also keep in mind the intricacies of \textit{chemical complexity} in reactive and/or functional molecular systems comprising numerous molecular transformation steps. Quantum computation will deliver only part of the results needed. Therefore, it will be necessary to integrate, at least partially, certain calculations offloaded to a quantum computer into existing software. Achieving this will require all quantum algorithms (also those for the calculation of molecular properties and forces) to mature to the level we are now approaching in single-point quantum energy evaluations. Work along these lines has already begun\cite{castaldo2022quantum, steudtner2023fault, castaldo2024differentiable, izsak2023quantum, huang2025fullqubit, gunther2025use, baker2026qdk}. Still, too little attention has been paid to algorithm development beyond (ground state) energies.

Beyond the scope of quantum chemistry, quantum variational algorithms for machine learning require numerous high-quality circuit evaluations to estimate gradients via sampling. Improving the gate operation speed in quantum computers would reduce algorithmic overhead, potentially extending potential quantum advantages to a broader range of problem sizes and instances—--not only in the asymptotic regime.

We conclude this analysis by pondering the development and operational cost of a quantum computer. Recent advances in real-time Field-Programmable Gate Array based error decoding routines~\cite{barber2025real} reinforce the emerging paradigm of quantum computing integrated with existing classical high-performance computing (HPC) centers~\cite{viviani2025assessing, delgado2025defining}. With current quantum hardware technology, quantum HPC centers will differ substantially from present-day classical facilities, particularly because the energy required for cooling quantum processors is expected to exceed the energy required for the computation itself, with cooling loads being highly sensitive to the computational architecture~\cite{martin2022energy}. 

A comprehensive assessment of the energy and economic cost of a quantum computer must account for four intertwined components. First, the logical cost concerns the algorithmic resources necessary to perform quantum computations. As thoroughly discussed in Ref.\,\citenum{jaschke2023quantum}, larger entanglement resources favor quantum computation over classical computation. This is at the basis of a study\,\cite{berger2021quantum} that estimated a five orders of magnitude energy gain over classical computation in Google’s quantum advantage landmark experiment\cite{arute2019quantum}. Nonetheless, accounting only for this energy cost would be a too simplistic approach. Such quantum machines incur an operational cost that encompasses the physical and thermodynamic resources required to run the processor\cite{fellous2023optimizing}. 

At this level, quantum thermodynamics\cite{campbell2025roadmap} becomes central since maintaining quantum coherence requires continuous entropy reduction, making the external control (i.e., qubit manipulation, physical interconnection, isolation systems, and readout infrastructure) energetically demanding. Hence, a strong emphasis must be put on improved refrigeration technologies, efficient error decoding, and optimized quantum control schemes. Increasing noise resilience requires additional algorithmic resources, such as larger and more efficient quantum error correction schemes, which in turn demand more physical qubits and more sophisticated classical control systems, thereby increasing classical energy consumption and hardware overhead. Third, we cannot neglect production costs including fabrication complexity, supply chains, and the use of critical raw materials necessary to build scalable quantum hardware. Moreover, end-of-life management raises questions concerning disposal and recyclability of cryogenic components, superconducting materials, rare elements, and electronic subsystems\cite{arora2024sustainable}.

Accordingly, no definitive quantitative answer exists regarding the full cost of quantum computing infrastructures and their potential energy advantage. Yet, we need to take into account the above constraints in a holistic framework to evaluate the long-term edge of quantum technologies, and ultimately, quantum computation at utility scale.

\section{Author information}
\textbf{Corresponding Author}

Markus Reiher - ORCID: 0000-0002-9508-1565 ; 

Email: markus.reiher@phys.chem.ethz.ch

\textbf{Authors} 

Davide Castaldo - ORCID: 0000-0001-8622-175X

\textbf{Author Contributions} \\
Both authors contributed to conceptualize the work, summarize the literature, design the figures and write the paper. Moreover both authors reviewed and edited the manuscript.


\begin{acknowledgement}
The authors acknowledge financial support from the Swiss National Science Foundation through Grant No. 200021\_219616 and through Grant No. 233975 (SPEED-Q Swiss National Postdoctoral Fellowship) and from the Novo Nordisk Foundation (Grant No. NNF20OC0059939 'Quantum for Life'). 
\end{acknowledgement}

\providecommand{\latin}[1]{#1}
\makeatletter
\providecommand{\doi}
  {\begingroup\let\do\@makeother\dospecials
  \catcode`\{=1 \catcode`\}=2 \doi@aux}
\providecommand{\doi@aux}[1]{\endgroup\texttt{#1}}
\makeatother
\providecommand*\mcitethebibliography{\thebibliography}
\csname @ifundefined\endcsname{endmcitethebibliography}
  {\let\endmcitethebibliography\endthebibliography}{}

\end{document}